\documentclass[10pt,a4paper]{article}
\usepackage{graphicx}
\usepackage{amsmath}
\usepackage{hyperref}
\usepackage{amssymb}
\hypersetup{colorlinks,linkcolor={blue},citecolor={blue},urlcolor={blue}}
\usepackage[version=4]{mhchem}
\usepackage{mathtools, nccmath}
\usepackage{setspace}
\usepackage{etoc}
\usepackage{cite}
\usepackage{float}
\usepackage{lipsum}
\usepackage{blindtext}
\usepackage[table,xcdraw]{xcolor}
\usepackage{multirow}
\usepackage{booktabs}
\usepackage{siunitx}
\usepackage{booktabs}
\usepackage{outline}
\usepackage{subcaption}
\usepackage{avant} 
\usepackage{multirow}
\usepackage[table,xcdraw]{xcolor}

\usepackage[a4paper, margin=1in, centering]{geometry}
\usepackage{array}
\usepackage{rotating}
\usepackage{epstopdf}
\usepackage{tikz}
\usepackage[compat=1.1.0]{tikz-feynman}
\usepackage{lscape}
\usepackage{subcaption}
\usepackage[toc,page]{appendix}
\usepackage{enumitem}
\usepackage{comment}
\newgeometry{vmargin={10mm,20mm},hmargin={19mm,19mm}}

\title{\textbf{\textcolor{blue}{Type-III Seesaw in Non-Holomorphic Modular Symmetry and Leptogenesis}}}

\author{{Priya}{\footnote{kashyappriya963@gmail.com}, Labh Singh\footnote{sainilabh5@gmail.com}, B. C. Chauhan\footnote{bcawake@hpcu.ac.in}}, and Surender Verma\footnote{s\_7verma@hpcu.ac.in}}

\date{\textit{Department of Physics and Astronomical Science\\ Central University of Himachal Pradesh, Dharamshala (HP) 176215, INDIA}}

\begin{document}
\maketitle  
\begin{abstract}
\noindent Recently, Qu and Ding, have proposed a formalism where modular invariance is extended to non-supersymmetric scenario considering Yukawa couplings as non-holomorphic functions of modules field $\tau$. Adopting this formalism in this work, we propose a Type-III seesaw model as a unified framework to explain lepton masses and mixing and baryogenesis via leptogenesis. $\chi^2$ analysis is performed to fit the neutrino oscillation data from NuFIT~6.0 leading to a normal hierarchical pattern of neutrino masses and constrained $CP$ phases. Furthermore, we analyze the generation of the observed baryon asymmetry of the Universe via thermal leptogenesis where the decays of the lightest fermion triplet $\Sigma_1$ into lepton-Higgs final states produce a $CP$ asymmetry $\varepsilon_{CP}$. The complex modules $\tau$ is responsible for the $CP$ asymmetry produced during leptogenesis. The washout processes dominated by gauge scatterings and inverse decays are studied through the full set of Boltzmann equations. The resulting $B-L$ asymmetry, $Y_{B-L}\sim 10^{-9}$ successfully reproduces the baryon-to-photon ratio demonstrating the model’s capability to link low-energy neutrino data with the baryogenesis.  The 
strong gauge-mediated washout of fermion triplets necessitates a leptogenesis scale of $\mathcal{O}(10^{12}\,\mathrm{GeV})$ ensuring compatibility with both the Davidson-Ibarra bound and the thermal history of the Universe. Future pursuits remain open to the exploration of novel avenues aimed at lowering the energy scale associated with leptogenesis.  

\end{abstract}

\section{Introduction}

The Standard Model (SM) of particle physics successfully describes the interactions of elementary particles, with the exception of gravity. However, several questions remain answered such as the origin and smallness of neutrino masses, the nature of dark matter, and the observed matter--antimatter asymmetry in the Universe. Experimental evidence from neutrino oscillation phenomena confirms that neutrinos are massive particles. 
The current values of neutrino oscillation parameters are reported in the latest global fit data~\cite{Esteban:2024eli}. Despite significant progress, several important aspects remain unresolved, including the hierarchy of neutrino masses, the octant of the atmospheric mixing angle ($\theta_{23}$), the mechanism responsible for generating neutrino masses, the fundamental nature of neutrinos (whether Dirac or Majorana), and the possible existence of $CP$ violation in the lepton sector.

\noindent The non-Abelian flavor symmetries are widely studied as a framework for understanding neutrino mass generation and mixing, with various discrete groups such as $S_3, A_4,S_4$, etc, commonly used to explain observed neutrino mixing patterns~\cite{Petcov:2017ggy, Ding:2024ozt, Ishimori:2010au,Chauhan:2023faf}. There conventional approaches have several key drawbacks, such as they often require a large number of flavon fields, whose specific vacuum alignments strongly influence the flavor structure of quarks and leptons making model predictions highly sensitive. Additional symmetries are usually needed to suppress unwanted terms in the Yukawa Lagrangian. Moreover, the flavor symmetry-breaking sector introduces many unknown parameters which reduces the minimality and predictability of the model. In order to overcome these limitations and to develop a lineage, it has been proposed that certain non-Abelian discrete symmetries can arise from superstring theory through the compactification of extra dimensions, leading to the modular invariance approach~\cite{Feruglio:2017spp, Feruglio:2019ybq, Ohki:2020bpo}. Unlike conventional models, modular symmetry frameworks do not require flavon fields, as the flavor symmetry can be broken solely by the vacuum expectation value ($vev$) of the modulus $\tau$ thus simplifying the flavor model construction. In recent years, modular symmetries have emerged as a compelling alternative, where Yukawa couplings are promoted to modular forms of a finite modular group $\Gamma_N$, and the $vev$ of the modulus $\tau$ plays the role of symmetry breaking. In particular, holomorphic modular symmetry has been implemented in supersymmetric (SUSY) contexts where the superpotential must be a holomorphic function of $\tau$~\cite{Feruglio:2017spp}. These models have demonstrated remarkable predictivity with successful constructions explaining lepton masses and mixing using a small number of input parameters~\cite{deAdelhartToorop:2011re}. However, holomorphic modular forms are severely constrained: the space of such forms is finite and limited to non-negative integral weights, which can restrict the richness of flavor textures, especially in non-SUSY or radiative mass-generation models. Alternatively, in the absence of low-energy supersymmetry—as is increasingly suggested by null results in SUSY searches—one is naturally led to consider a non-SUSY realization of modular symmetry\cite{ParticleDataGroup:2024cfk,Cremades:2004wa,Almumin:2021fbk}. In this context, the recently proposed non-holomorphic modular symmetry based on harmonic or polyharmonic Maaß forms provides a powerful generalization~\cite{Qu:2024rns}. These forms are modular-invariant but need not be holomorphic satisfying Laplace-type differential equations instead. This broader function space includes both holomorphic and non-holomorphic parts and accommodates modular forms of both positive and negative weights allowing for more flexible and phenomenologically viable Yukawa structures. Importantly, the modular invariance continues to constrain the form of the Yukawa couplings maintaining predictivity while overcoming the limitations of the holomorphic approach. Non-holomorphic modular symmetries have been explored in various theoretical frameworks including the Type II seesaw mechanism~\cite{Nomura:2024atp}, the scotogenic model~\cite{Nomura:2024vzw}, the Zee–Babu model~\cite{Kobayashi:2025hnc}, and other extensions of the SM that address neutrino masses and flavor structure~\cite{Ding:2024inn, Li:2024svh,Loualidi:2025tgw,Zhang:2025dsa,Nomura:2025raf,Nomura:2025ovm, Kumar:2025bfe, Nomura:2025bph, Nomura:2024ctl, Okada:2025jjo,Nomura:2024nwh}.

\noindent In this work, we propose a novel realization of the Type-III seesaw mechanism~\cite{Foot:1988aq} under the framework of non-holomorphic modular symmetry employing polyharmonic Maaß forms of a finite modular group $\Gamma_3$. We show that the Yukawa couplings and mass matrices can be organized as multiplets of $\Gamma_3$, and demonstrate that the presence of negative weight modular forms enhances the model’s capacity to reproduce the observed lepton mixing patterns and leptogenic $CP$ asymmetry. This framework also provides new avenues to connect low-energy neutrino phenomenology with high-energy UV completions, including string theory constructions where non-holomorphic automorphic forms naturally arise. Further, we have also explored the baryogenesis through leptogenesis by the decay of lightest fermion triplet. It is to be noted that complex modulus($\tau$) is the only source of $CP$ violation in the model. In the present work, we consider the thermal leptogenesis based on non-holomorphic modular weights and the mass of the fermion triplet is of the order of $10^{12}$ GeV which satisfies the Davidson-Ibarra bound on decaying lightest right-handed neutrino, in order to explain the baryon asymmetry of the universe (BAU)~\cite{Davidson:2002qv}.

\noindent The structure of the paper is as follows: Section \ref{section2} provides a brief overview of non-holomorphic modular symmetry. Section \ref{section3} presents the proposed model on the Type-III seesaw mechanism with non-holomorphic modular symmetry, followed by the numerical analysis in Section \ref{section4}. Leptogenesis is discussed in Section \ref{section5}, and conclusions are summarized in Section \ref{section6}.

\section{Non-Holomorphic Modular Symmetry}
\label{section2}
\noindent Modular symmetry arises as a geometric symmetry of a two-dimensional torus. The 2-dimensional torus can be constructed by the division of 2-D Euclidean space($\mathbb{R}^2$) by the lattice $\Lambda$, i.e., $T^2 = \frac{\mathbb{R}^2}{\Lambda}$.  Alternatively, this can be described on the 1-D complex plane, where the lattice is generated by two basis vectors, and their ratio defines the complex structure of the torus~\cite{Ishimori:2010au,Chauhan:2023faf}. This ratio is denoted by the complex modulus $\tau = x+ iy$ and given as 

\begin{equation}
    \tau' \rightarrow \tau = \gamma \tau = \begin{pmatrix}
        a\tau & b \\
        c\tau & d  
    \end{pmatrix}.
\end{equation}
The integers $a, b, c, d$ satisfy the condition $ad - bc = 1$, defining the group $SL(2,\mathbb{Z})$ of modular transformations. This group captures the modular symmetry of the torus. Finite modular subgroups $\Gamma_N$ (for $N = 2, 3, 4, 5$) are isomorphic to well-known finite groups $ \Gamma_2 \simeq S_3,\Gamma_3 \simeq A_4, \Gamma_4 \simeq S_4, 
\Gamma_5 \simeq A_5$ respectively \cite{Ishimori:2012zz}.

\noindent Modular invariance provides a well-motivated framework for constructing flavor models that aim to explain the hierarchical structure of fermion masses and mixings. In particular, $\mathcal{N} = 1$ SUSY models that incorporate modular symmetry, especially within a bottom-up approach, have demonstrated notable success in reproducing realistic lepton mass spectra and mixing patterns with a minimal number of free parameters \cite{Feruglio:2017spp}. A fundamental characteristic of these models is that the matter fields appearing in the superpotential are required to transform under representations of the modular group, either $SL(2, \mathbb{Z})$ or its projective form $PSL(2, \mathbb{Z})$. This requirement imposes that the Yukawa couplings must be expressed as holomorphic functions of a complex modulus $\tau$, which resides in the upper half-plane of the complex domain.

\noindent In the modular framework, SUSY ensures the holomorphic structure of Yukawa coupling as modular form in the superpotential. However, constraints from flavor physics, high-precision electroweak measurements, and astrophysical observations place strong limits on the allowed SUSY parameter space. So far, no experimental evidence for low-energy SUSY has been found, and it remains uncertain whether SUSY exists in nature or how it might be realized. This has led to growing interest in exploring modular invariance without assuming supersymmetry. Many authors have been explored modular symmetry in $\Gamma(3) = A_4$ group \cite{Mishra:2022egy ,Kashav:2021zir,Kashav:2022kpk,Singh:2024imk, Kashav:2024lkr, Kumar:2023moh, Behera:2020lpd, Behera:2020sfe}.
It has been proposed that the presence of low-energy SUSY is not required for Yukawa couplings to exhibit modular form behavior. The amount of SUSY that remains at low energies depends on the specific geometry of the compactification. Additionally, a non-SUSY realization of modular flavor symmetry has been introduced using the theory of automorphic forms. In this framework, the usual requirement of holomorphicity is replaced by Laplacian condition. For a single complex modulus, automorphic forms correspond to harmonic Maaß forms, which are non-holomorphic modular functions satisfying the Laplace equation. While modular flavor symmetry is often considered within a SUSY context, the automorphic forms framework enables a non-SUSY realization of modular flavor symmetry, as recently proposed in \cite{Qu:2024rns}. The polyharmonic Maaß form is the automorphic form associated with the single modulus  $\tau$ . In the case with the polyharmonic Maaß form, the level N and the finite modular group $\Gamma'_N$ (or $ \Gamma_N$) are kept fixed. The generic matter fields are denoted by $ \psi_i$ and $\psi^c_i $, and their transformations under the modular group are specified by the modular weights $ -k_\psi $, $ -k_{\psi^c} $, and the irreducible representations $ \rho_\psi $, $ \rho_{\psi^c} $ of the finite modular group $ \Gamma'_N $ (or $ \Gamma_N $), i.e.,

\begin{equation}
   \tau \to \gamma \tau = \frac{a\tau + b}{c\tau + d}, \quad \gamma = \begin{pmatrix} a & b \\ c & d \end{pmatrix} \in SL(2, \mathbf{Z}), 
\end{equation}

\begin{equation}
  \psi_i(x) \to (c\tau + d)^{-k_\psi} [\rho_\psi(\gamma)]_{ij} \psi_j(x), \quad \psi^c_i(x) \to (c\tau + d)^{-k_{\psi^c}} [\rho_{\psi^c}(\gamma)]_{ij} \psi^c_j(x).  
\end{equation}

\noindent Our primary focus is on the fermion mass terms generated by the Yukawa interactions of Dirac fermions. The Yukawa interaction, invariant under modular symmetry, can be expressed as follows

\begin{equation}
    \mathcal{L}_Y = Y(k_Y)(\tau) \psi^c \psi H + \text{h.c.},
\end{equation}
\noindent Here, the fermion fields are expressed using two-component spinor notation, and $H$ denotes the Higgs field (or its complex conjugate). The Higgs field transforms under the modular symmetry as follows

\begin{equation}
  H(x) \to (c\tau + d)^{-k_H} \rho_H(\gamma) H(x).  
\end{equation}
\noindent The modular invariant Lagrangian is constructed from matter fields and polyharmonic Maaß forms of level $N$. The polyharmonic Maaß form $Y(\tau)$, with weight $k$ at level $N$, depends on the complex modulus $\tau$ and transforms under modular symmetry according to

\begin{equation}
    Y(\gamma\tau) = (c\tau + d)^{k_Y} Y(\tau) \quad \gamma = \begin{pmatrix}
        a & b \\
        c & d 
    \end{pmatrix} \in \Gamma(N).
\end{equation}
Moreover, these forms must satisfy the Laplacian condition as well as appropriate growth conditions, which are given by
\begin{equation}
    -4y^2 \frac{\partial}{\partial \tau} \frac{\partial}{\partial \bar{\tau}} + 2ik_Yy \frac{\partial}{\partial \bar{\tau}} Y(\tau) = 0,
\end{equation}
\begin{equation}
   Y(\tau) = O(y^{\alpha}) \quad \text{as } y \to +\infty, \text{ uniformly in } x. 
\end{equation}
Each term in the Yukawa interaction $L_Y$ must be invariant under the finite modular group $\Gamma'_N $ (or $ \Gamma_N $), and its total modular weight must vanish. Therefore, $ k_Y $ and $ \rho_Y $ must satisfy the following conditions

\begin{equation}
    k_Y = k_{\psi^c} + k_\psi + k_H,
\end{equation}
\begin{equation}
    \rho_Y \otimes \rho_{\psi^c} \otimes \rho_\psi \otimes \rho_H \supset 1,
\end{equation}
Here, $1$ denotes the trivial singlet representation of $\Gamma'_N$ (or $\Gamma_N$). For Majorana fermions, represented by $\psi^c$, the associated mass terms are given by
\begin{equation}
L_M = Y(k_Y)(\tau)\, \psi^c \psi^c + \text{h.c.}
\end{equation}
Similarly, the conditions for modular invariance are
\begin{equation}
k_Y = 2 k_{\psi^c}, \quad \rho_Y \otimes \rho_{\psi^c} \otimes \rho_{\psi^c} \supset 1.
\end{equation}

\begin{table}[t]
    \centering
    \begin{tabular}{l l}
    \hline
    Weight $k_Y$ & Polyharmonic Maaß forms $Y_r^{(k_Y)}$ \\
    \hline
    $k_Y = -4$ & $Y_1^{(-4)}$, \quad $Y_3^{(-4)}$ \\
    $k_Y = -2$ & $Y_1^{(-2)}$, \quad $Y_3^{(-2)}$ \\
    $k_Y = 0$  & $Y_1^{(0)}$, \quad $Y_3^{(0)}$ \\
    $k_Y = 2$  & $Y_1^{(2)}$, \quad $Y_3^{(2)}$ \\
    $k_Y = 4$  & $Y_1^{(4)}$, \quad $Y_1^{(4)}$, \quad $Y_3^{(4)}$ \\
    $k_Y = 6$  & $Y_1^{(6)}$, \quad $Y_{31}^{(6)}$, \quad $Y_{3I}^{(6)}$ \\
    \hline
    \end{tabular}
    \caption{Summary of polyharmonic Maaß forms of weight $k_Y = -4, -2, 0, 2, 4, 6$ at level $N = 3$, the subscript $r$ denote the transformation property under $A_4$ modular symmetry. Here $Y_{31}^{(6)}$ and $Y_{3I}^{(6)}$ stand for two independent weight 6 modular forms transforming as a triplet \textbf{3} of $A_4$.}
\end{table}

\section{Model and Formalism}
\label{section3}
\noindent In this study, we extend the SM by introducing a fermion triplet which gives rise to the Type-III seesaw mechanism for neutrino mass generation. The left-handed (LH) lepton doublets are assigned to transform as a triplet under the modular $A_4$ symmetry with modular weight 0. The right-handed (RH) charged leptons denoted as $E_1$, $E_2$, and $E_3$, transform as $1$, $1''$, and $1'$, respectively, under $A_4$ with corresponding modular weights of $-2$, $0$, and $0$. The Higgs field remains invariant under the $A_4$ symmetry and carries zero modular weight. We consider the modular weights in the range $ -4 \leq k_I \leq 0$, since larger weights typically introduce more free parameters, which can reduce the predictability of model. The charge assignments under modular group $A_4$ and modular weights for our model are shown in Table 
\ref{tab:charges}. The modular invariant Lagrangian is given as
\begin{equation}
\begin{split}
    -\mathcal{L} =\; & \alpha\, \bar{L}\, H\, E_1\, Y_3^{(-2)} + \beta\, \bar{L}\,  H\, E_2\, Y_3^{(0)} 
    + \gamma\,\bar{L}\, H\,  E_3\, Y_3^{(0)} \\[5pt]
    & + g_1\, \bar{L}\, \Sigma_i\, H\,  Y_1^{(-2)} + g_2\, ( \bar{L}\,\Sigma_i\, H\,  Y_3^{(-2)})_{\text{sym}} 
    + g_2'\, (\bar{L}\, \Sigma_i\, H\,  Y_3^{(-2)})_{\text{asym}} \\[5pt]
    & + M_0\, \text{Tr}[\Sigma_i\, \Sigma_i]\, Y_1^{(-4)} + M'\, \text{Tr}
    [\Sigma_i\, \Sigma_i]\, Y_3^{(-4)} + \text{h.c.}.
\end{split}
\end{equation}

\noindent where, the phases of the couplings $\alpha$, $\beta$, $\gamma$, $g_1$, $g_2$, and $g_2'$ can be absorbed into the right-handed lepton fields. Consequently, the $vev$ of $\tau$ becomes the unique source of $CP$ violation. The modular invariant Lagrangian for the charged lepton sector is given as

\begin{align}
    -L_e &= \alpha \bar{L} H E_1  Y_3^{(-2)} +  \beta \bar{L} H   E_2 Y_3^{(0)} + \gamma \bar{L}  H E_3 Y_3^{(0)} + \text{h.c.},
\end{align}
which corresponds to a diagonal charged lepton mass matrix and is given as 
\begin{equation}
    M_L = \text{Diag}(\alpha,\beta,\gamma) 
    \begin{pmatrix}
        Y_{31}^{(-2)} & Y_{33}^{(-2)} & Y_{32}^{(-2)} \\
        Y_{32}^{(0)} & Y_{31}^{(0)} & Y_{33}^{(0)} \\
        Y_{33}^{(0)} & Y_{32}^{(0)} & Y_{31}^{(0)}
    \end{pmatrix}v,
\end{equation}
\noindent where $v$ is the $vev$ of the Higgs field. The charged lepton mass matrix $ M_L$ contains three real parameters $ \alpha, \beta$ and $\gamma $ which can be suitably chosen to reproduce the observed charged lepton masses. The diagonalization of $M_L$ is given as $U_L^\dagger M_L^\dagger M_L U_L = \text{Diag}{(|m_e|^2, |m_\mu|^2, |m_\tau|^2)}$. The modular invariant Lagrangian for the Dirac term is given as
\begin{equation}
    -L_D = g_1 \bar{L} \Sigma_i H Y_1^{(-2)}+  g_2 (\bar{L} \Sigma_i H Y_3^{(-2})_{\text{sym}} + g_2'( \bar{L} \Sigma_i H Y_3^{(-2)})_{\text{asym}}+ \text{h.c.}.
\end{equation}
In our model, $L$, $\Sigma_i$, and Yukawa coupling are triplet and at weight 2 and level 3, the space of polyharmonic Maaß forms includes non-holomorphic functions as the modified Eisenstein series $ E^{(b)}_2(\tau) $, which transforms as a trivial singlet under the $ A_4 $ symmetry. The tensor product 3 $\otimes$ 3 is decomposed into symmetric and antisymmetric parts, and the corresponding Dirac mass matrix is given as 

\begin{equation}\label{mdyukawa}
    M_D =  v \begin{pmatrix}
        g_1 Y_1^{(-2)} + 2 \frac{g_2}{3} Y_{31}^{(-2)} &   (\frac{g_2'}{2}-\frac{g_2}{3}) Y_{33}^{(-2)}&  -(\frac{g_2'}{2}+\frac{g_2}{3}) Y_{32}^{(-2)} \\
         -(\frac{g_2'}{2}+\frac{g_2}{3})Y_{33}^{(-2)} & 2 \frac{g_2}{3} Y_{32}^{(-2)} & g_1 Y_1^{(-2)}+(\frac{g_2'}{2}-\frac{g_2}{3}) Y_{31}^{(-2)} \\
         (\frac{g_2'}{2}-\frac{g_2}{3})Y_{32}^{(-2)} & g_1 Y_1^{(-2)}- (\frac{g_2'}{2}+\frac{g_2}{3}) Y_{31}^{(-2)} & 2 \frac{g_2}{3} Y_{33}^{(-2)}
    \end{pmatrix}.
\end{equation}

\begin{table}[t]
\centering
\begin{tabular}{ccccccc}
\toprule
 & $\bar{L}$ & $E_1$&$E_2$&$E_3$ & $\Sigma_i$ & $H$   \\
\midrule
$SU(2)$ & $2$ & $1$ &$1$ &$1$ & $3$ & $2$  \\
$A_4$ & $3$ & $1$ & $1''$& $1'$ & $3$ & $1$   \\
$k_I$ & $0$ & $-2$&0&$0$ & $-2$ & $0$   \\
\bottomrule
\end{tabular}
\caption{Field content and charge assignments of Type-III seesaw mechanism under $A_4$ group and modular weights.}
\label{tab:charges}
\end{table}

\noindent The invariant modular Lagrangian for the Majorana term is given as 
\begin{equation}
    -L_\Sigma = 
M_0  Tr[\Sigma_i \Sigma_i] Y_1^{(-4)} + M'  Tr[\Sigma_i \Sigma_i] Y_3^{(-4)}  + \text{h.c.},
\label{Ln}
\end{equation}
where \(\Sigma_i\) with i = $1,2,3$ can be presented in \(SU(2)\) basis as
\begin{equation}
    \Sigma_i = \begin{pmatrix}
        \frac{\Sigma_i^0}{\sqrt{2}} & \Sigma_i^+ \\
        \Sigma_i^- & -\frac{\Sigma_i^0}{\sqrt{2}}
    \end{pmatrix}.
\end{equation}
\noindent The right-handed Majorana neutrino mass matrix is given as

\begin{equation}
    M_\Sigma =  \begin{pmatrix}
       M_0 Y_1^{(-4)}+ 2 M' Y_{31}^{(-4)} &   -M'Y_{33}^{(-4)}& - M' Y_{32}^{(-4)} \\
         - M' Y_{33}^{(-4)} & 2M' Y_{32}^{(-4)} &  M_0 Y_1^{(-4)}-M'Y_{31}^{(-4)} \\
         -M'Y_{32}^{(-4)} &  M_0 Y_1^{(-4)}-M'Y_{31}^{(-4)} & 2 M'Y_{33}^{(-4)}
    \end{pmatrix}.
\end{equation}
The mass matrix $M_\Sigma$, can be diagonalized by using the unitary matrix, such that, $U_R$ as $U_R^TM_\Sigma U_R = \text{diag}(M_{\Sigma_1},M_{\Sigma_2},M_{\Sigma_3})$.
\noindent Finally, the active neutrino mass matrix is obtained by the Type-III seesaw as follows
\begin{equation}
M_\nu = -M_D M_\Sigma^{-1} M_D^T.
\label{neutrinomassmatrix}
\end{equation}
The neutrino mass matrix given by Eqn. (\ref{neutrinomassmatrix}) is diagonalized using the relation $U_\nu^T M_\nu U_\nu = \text{diag}(m_{\nu_1}, m_{\nu_2}, m_{\nu_3})$. The mixing matrix $U_{PMNS}$ = $U_L^\dagger U_\nu$, since the charged lepton mass matrix is not diagonal in the flavor basis. Now, the mixing angle can be extracted from $U_{PMNS}$ as 

\begin{equation}
    \sin^2 \theta_{13} = |U_{13}|^2, \quad
    \sin^2 \theta_{12} = \frac{|U_{12}|^2}{1 - |U_{13}|^2}, \quad
    \sin^2 \theta_{23} = \frac{|U_{23}|^2}{1 - |U_{13}|^2},
\end{equation}
where $U_{mn}$ are elements of $U_{PMNS}$. The Dirac $CP$ violating phase $\delta_{CP}$ can be determined from the PMNS matrix elements through the Jarlskog invariant defined as
\begin{equation}
    J_{CP} = \text{Im}\left[U_{11} U_{22} U^{*}_{12} U^{*}_{21}\right] = s_{23} c_{23} s_{12} c_{12} s_{13} c_{13}^2 \sin\delta_{CP},
\end{equation}
where \( s_{ij} = \sin\theta_{ij} \) and \( c_{ij} = \cos\theta_{ij} \). In addition to $\delta_{CP}$, the Majorana $CP$ phases can be investigated using the PMNS matrix elements as
\begin{equation}
    I_1 = \text{Im}\left[U_{11}^* U_{12}\right] = c_{12} s_{12} c_{13}^2 \sin\left(\frac{\alpha_{21}}{2}\right),
    \label{I1}
\end{equation}
\begin{equation}
    I_2 = \text{Im}\left[U_{11}^* U_{13}\right] = c_{12} s_{13} c_{13} \sin\left(\frac{\alpha_{31}}{2} - \delta_{CP}\right).
    \label{I2}
\end{equation}

\noindent The effective Majorana mass($m_{\beta \beta}$) is given as 
\begin{equation}
    m_{\beta \beta} = \left| 
m_{1} \cos^2\theta_{12} \cos^2\theta_{13} 
+ m_{2} \sin^2\theta_{12} \cos^2\theta_{13} e^{i\alpha_{21}} 
+ m_{3} \sin^2\theta_{13} e^{i(\alpha_{31} - 2\delta_{\text{CP}})} 
\right|
\end{equation}
In the following section, we will numerically diagonalize the neutrino mass matrix in Eqn. (\ref{neutrinomassmatrix}) and elucidate predictions for the neutrino observables. 

\section{Numerical Analysis and Discussion}
\label{section4}
\noindent In this section, we have performed numerical analysis and test the model against the experimental data shown in Table \ref{tab:neutrino_data}. In order to  ascertain allowed parameter space we define $\chi^2$ function as
\begin{equation}
\centering
\chi^2 = \sum_{i = 1}^7 \left( \frac{P_i - P_i^0}{\sigma_i} \right)^2,
     \label{}
\end{equation}
where $P_i$ represents the observables predicted by the model, $P_i^0$ represents the central value and $\sigma_i$ represents the error corresponding to the 1$\sigma$ level. We have sampled the seven observables in this study, which are three mixing angles, two mass-squared differences ($\Delta m^2_{21}$ and $\Delta m^2_{31}$), and two lepton mass ratios($m_e/m_\mu$ and $m_\mu/m_\tau$). The experimental values used in our analysis are listed in Table \ref{tab:neutrino_data}. 

\begin{figure}[t]
\begin{subfigure}{0.33\textwidth}
  \includegraphics[width=1\linewidth]{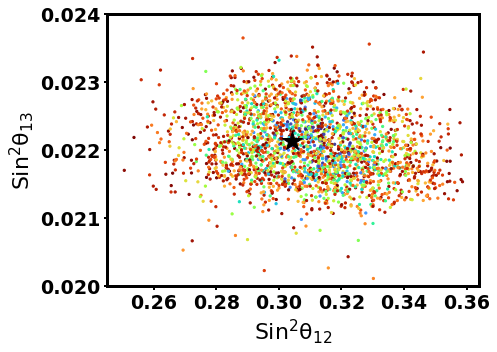}
  \caption{}
  \label{Fig1}
\end{subfigure}%
\hfill
\begin{subfigure}{0.33\textwidth}
  
  \includegraphics[width=1\linewidth]{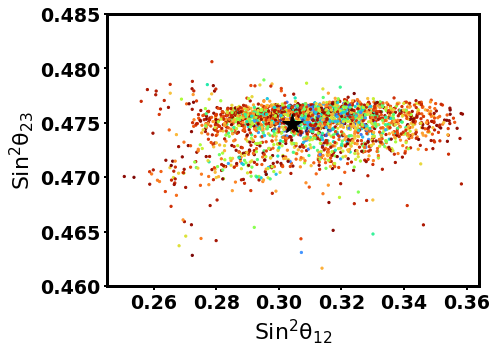}
  \caption{}
  \label{Fig2}
\end{subfigure}%
\hfill
\begin{subfigure}{0.33\textwidth}
  
  \includegraphics[width=1\linewidth]{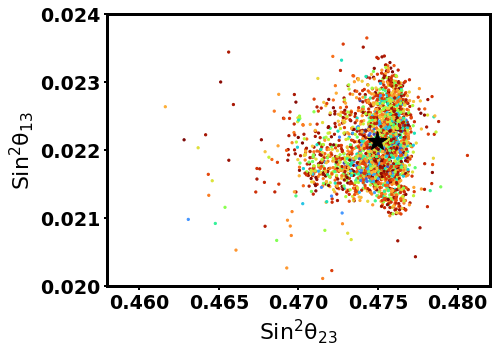}
  \caption{}
  \label{Fig3}
\end{subfigure}%
\hfill
\begin{subfigure}{0.32\textwidth}
  
  \includegraphics[width=1\linewidth]{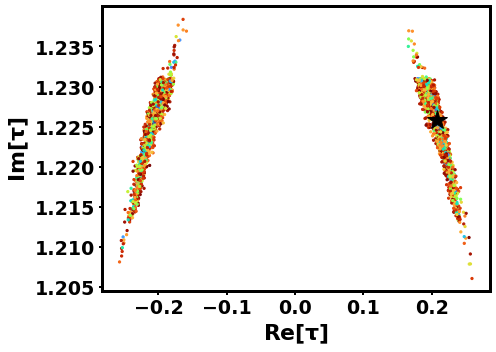}
  \caption{}
  \label{Fig4}
  \label{Fig1NH-d}
\end{subfigure}
\hfill
\begin{subfigure}{0.32\textwidth}
    \includegraphics[width=1\linewidth]{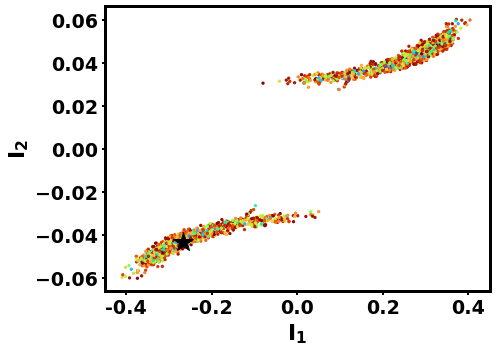}
    \caption{}
    \label{Fig6}
\end{subfigure}
\hfill
\begin{subfigure}{0.33\textwidth}
  
  \includegraphics[width=1\linewidth]{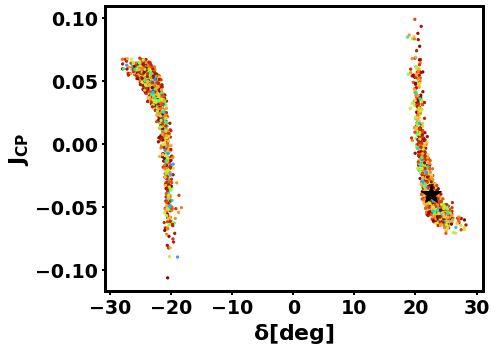}
  \caption{}
  \label{Fig8}
\end{subfigure}%
\hfill
\begin{subfigure}{0.33\textwidth}
  
  \includegraphics[width=1\linewidth]{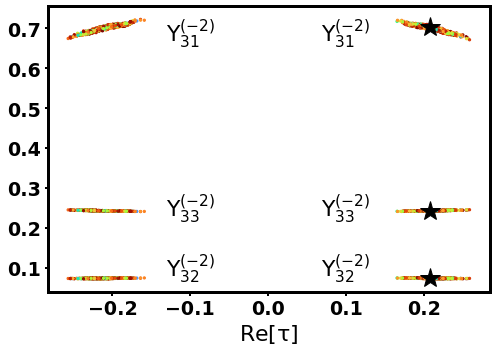}
  \caption{}
  \label{Fig9}
\end{subfigure}%
\hfill
\begin{subfigure}{0.33\textwidth}

  \includegraphics[width=1\linewidth]{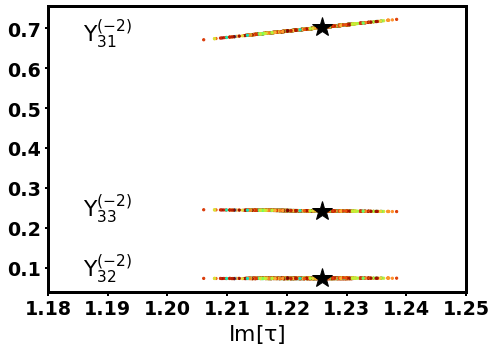}
  \caption{}
  \label{Fig10}
\end{subfigure}
\hfill
    \begin{subfigure}{0.33\textwidth}
  
  \includegraphics[width=1\linewidth]{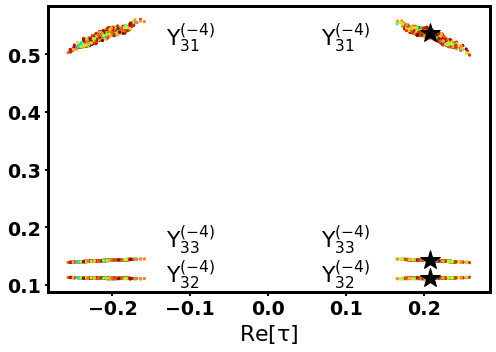}
  \caption{}
  \label{Fig11}
\end{subfigure}%
\hfill
\begin{subfigure}{0.33\textwidth}
  
  \includegraphics[width=1\linewidth]{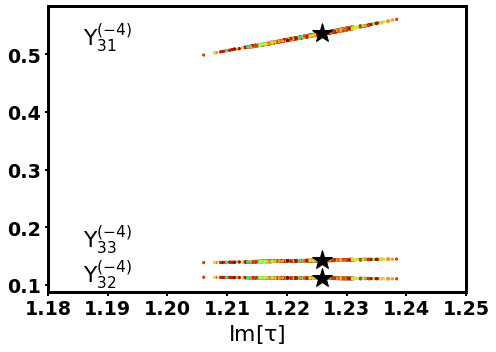}
  \caption{}
  \label{Fig12}
\end{subfigure}
\hfill
    \begin{subfigure}{0.33\textwidth}
  
  \includegraphics[width=1\linewidth]{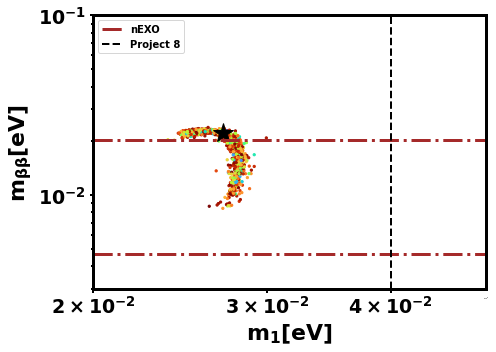}
  \caption{}
  \label{Fig13}
\end{subfigure}%
\hfill
\begin{subfigure}{0.33\textwidth}
  
  \includegraphics[width=1\linewidth]{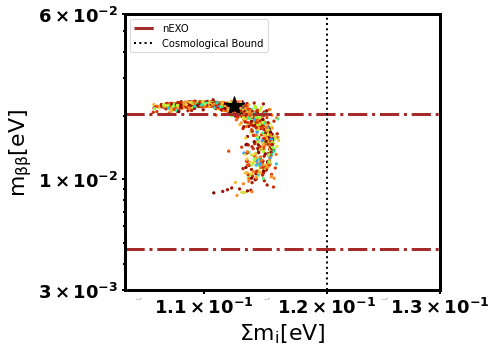}
  \caption{}
  \label{Fig14}
\end{subfigure}
\centering
\begin{subfigure}{0.5\textwidth}
    \includegraphics[width=1\linewidth]{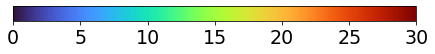} 
  \label{Fig15}
\end{subfigure}
\caption{Correlation plots are presented among neutrino observables, including leptonic mixing angles, atmospheric mixing angle, Dirac and Majorana $CP$ phases, Jarlskog invariant, and modular Yukawa couplings (weights –2 and –4) with respect to the real and imaginary parts of the complex modulus. We also show the effective Majorana mass versus the lightest neutrino mass and the sum of neutrino masses, with experimental bounds from nEXO (horizontal lines), Project 8 (vertical line). The solid-star represents best-fit values corresponding to $\chi^2_{min}=1.03$.}
\label{fig}
\end{figure}

\begin{table}[t]
\small
    \centering
    \renewcommand{\arraystretch}{1} 
    \begin{tabular}{l l l l l} 
        \hline
        Parameter & best-fit$\pm1\sigma$ range (NH) & best-fit$\pm1\sigma$ range (IH) & $3\sigma$ range (NH) & $3\sigma$ range (IH) \\
        \hline
        $\sin^2\theta_{12}$ & $0.308^{+0.012}_{-0.011}$ & $0.308^{+0.012}_{-0.011}$ & $0.275 - 0.345$ & $0.275 - 0.345$ \\
        $\sin^2\theta_{23}$ & $0.470^{+0.017}_{-0.013}$ & $0.562^{+0.012}_{-0.015}$ & $0.435 - 0.585$ & $0.410 - 0.623$ \\
        $\sin^2\theta_{13}$ & $0.02215^{+0.00056}_{-0.00058}$ & $0.02224^{+0.00056}_{-0.00057}$ & $0.02023 - 0.02388$ & $0.02053 - 0.02397$ \\
        $\Delta m^2_{3l} \times 10^{-3} \text{eV}^2$ & $2.513^{+0.021}_{-0.019}$ & $-2.510^{+0.024}_{-0.025}$ & $2.463 - 2.606$ & $-2.584 - -2.438$ \\
        $\Delta m^2_{21} \times 10^{-5} \text{eV}^2$ & $7.49^{+0.19}_{-0.19}$ & $7.49^{+0.19}_{-0.19}$ & $6.92 - 8.05$ & $6.92 - 8.05$ \\
        
        $m_e/m_\mu$ & $0.004737$ & $0.004737$ & -- & -- \\
        $m_\mu/m_\tau$ & $0.058823 $ & $0.058823$ & -- & -- \\
        \hline
    \end{tabular}
    \caption{The neutrino oscillation data used in the numerical analysis taken from NuFIT 6.0~\cite{Esteban:2024eli}. The central values of the charged lepton mass ratios are taken from Ref. \cite{Xing:2007fb}. During the scan of the model parameter space, we have used the 3 $\sigma$ range of quantities.}
    \label{tab:neutrino_data}
\end{table}
\noindent Fig.~\ref{fig}(\subref{Fig1}),~\ref{fig}(\subref{Fig2}),~\ref{fig}(\subref{Fig3}) shows how well the model describes the mixing angles in the case of NH with the minimum value of $\chi^2_{min}$ =  1.03. The predictions of the model have been depicted as correlation plots obtained for $\chi^2 \leq 30$. The solid-star in the figures represents best-fit values corresponding to $\chi^2_{min}=1.03$. The best-fit values for the neutrino observables, i.e., \(\sin^2\theta_{13}, \sin^2\theta_{12}, \sin^2\theta_{23},  \Delta m^2_{21}, \Delta m^2_{\text{31}}\)
are in the 1$\sigma$ range of experimental values as shown in Table \ref{tab:neutrino_params}. The atmospheric mixing angle ($\sin^2\theta_{23}$) clearly favors the first octant, with the range ($0.46-0.48$) at 3$\sigma$, which can be tested at the NO$\nu$A~\cite{2852068}, T2K experiment~\cite{Carabadjac:2024smm} DUNE~\cite{DUNE:2020fgq} and Hyper-Kamiokande experiments~\cite{Hyper-Kamiokande:2018ofw}. Fig. \ref{fig}(\subref{Fig4}) shows the allowed parameter space for the $\tau$ in the complex plane consistent with experimental data on lepton masses and mixing. It can be seen from Fig. \ref{fig}(\subref{Fig4}) that complex modules $\tau$ lie in its fundamental domain $D = \left\{ \tau \in \mathbb{C} \,\middle|\, \operatorname{Im} \tau > 0,\ |\operatorname{Re} \tau| \leq \frac{1}{2},\ |\tau| \geq 1 \right\}$ with best-fit corresponding to $\chi^2_{min}$ = 1.03 is $0.20+1.22i$. The Dirac-type $CP$ violating phase $\delta_{CP}$ is predicted to be non vanishing with its best-fit value corresponding to $\chi^2_{min}$ is  $22.5^o$ This is important in the sense that oscillation experiments~\cite{2852068, Carabadjac:2024smm, DUNE:2020fgq, Hyper-Kamiokande:2018ofw} aiming at measuring $\delta_{CP}$ may have a say about the viability of the model. The information about Majorana type $CP$ phases ($\alpha_{21}$ and $\alpha_{31}$) contained in $CP$ invariants ($I_1$ and $I_2$) defined in Eqn. (\ref{I1}) and (\ref{I2}) are shown as correlation plot in Fig. \ref{fig} (\subref{Fig6}). Fig.~\ref{fig}(\subref{Fig8}) shows the allowed parameter space for Jarlskog invariant $J_{CP}$ and $\delta$. Fig.~\ref{fig}(\subref{Fig9}), \ref{fig}(\subref{Fig10}) represents the allowed range of Yukawa coupling of modular weight $-2$ as a function of the real and imaginary part of the complex modulus $\tau$.

\begin{figure}
    \centering
    \includegraphics[width=0.5\linewidth]{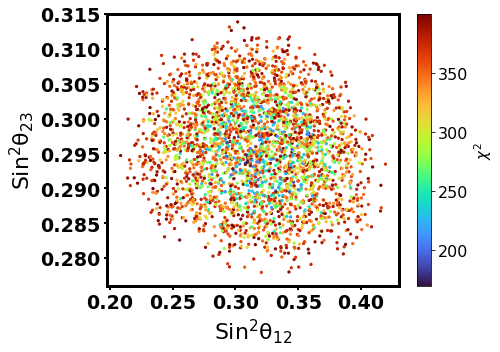}
    \caption{Correlation among the mixing angle $\sin^2\theta_{23}$ and $\sin^2\theta_{12}$ for the inverted hierarchy with the minimum value of $\chi^2_{min}=169.7$.}
    \label{fig:ih}
\end{figure}

\begin{table}[t]
    \centering
    \begin{tabular}{c c c c c c c c}
        \toprule
        $\sin^2\theta_{12}$ & $\sin^2\theta_{23}$ & $\sin^2\theta_{13}$ & $\delta_{CP} (^\circ)$ & $\Delta m^2_{21} (\text{eV}^2)$ & $\Delta m^2_{31} (\text{eV}^2)$ &\(\text{Re}[\tau]\) & \(\text{Im}[\tau]\) \\
        \midrule
        0.30 & 0.47 & 0.022 & $22.5^o$ & $7.5 \times 10^{-5}$ & $2.5 \times 10^{-3}$ & 0.20 & 1.22 \\
        \bottomrule
    \end{tabular}
    \caption{The best-fit values for the neutrino oscillation parameters obtained from the $\chi^2$ analysis correspond to a minimum $\chi^2$ value of 1.03 for NH.}
    \label{tab:neutrino_params}
\end{table}
\noindent Fig. \ref{fig}(\subref{Fig11}), \ref{fig}(\subref{Fig12}) represents the allowed range of Yukawa coupling of modular weight $-2$ as a function of the real and imaginary parts of the complex modulus $\tau$. \noindent Fig. \ref{fig}(\subref{Fig13}) represents the variation of the effective Majorana as a function of the lightest neutrino mass ($m_1$). It is evident from the plot that effective Majorana mass is in the sensitivity range with the proposed nEXO experiment \cite{nEXO:2021ujk} and $m_1$ also shows the compatibility with the projected sensitivities of KATRIN experiment ($0.2$ eV) \cite{KATRIN:2022ayy} and PROJECT 8 experiment \cite{Project8:2022wqh}. Fig. \ref{fig}(\subref{Fig14}) shows the effective Majorana mass with respect to the sum of three active neutrino masses ($\Sigma m_i$). It is observed that $\Sigma m_i$ is consistent with cosmological bound ($<0.12$ eV) at 95\% confidence level (CL) \cite{Planck:2018vyg}. Furthermore, we have investigated the parameter space for the inverted neutrino mass hierarchy (IH). The analysis excludes the IH scenario as illustrated in Fig.~\ref{fig:ih}. The figure depicts the correlation between the mixing angles $\sin^2\theta_{23}$ and $\sin^2\theta_{12}$, demonstrating that $\sin^2\theta_{23}$ lies outside the $3\sigma$ range of the global data thereby confirming the incompatibility of IH within the model.
\section{Leptogenesis}\label{section5}
\noindent In the Type-III seesaw model with non-holomorphic modular symmetry the hyperchargeless fermion triplet ($\Sigma_i$) interacts with the SM electroweak sector through gauge interactions. These interactions enable $\Sigma_i$ to participate in scattering with SM gauge bosons and annihilation into fermions and Higgs boson. Such processes influence the thermal history of $\Sigma_i$ affecting its decoupling temperature, number density evolution, and consequently, the efficiency of leptogenesis. The lepton asymmetry is generated through the $CP$-violating decays of the lightest triplet fermion ($\Sigma_1$), $\Sigma_1 \to L_\alpha H^\dagger$, arising from interference between the tree-level and one-loop (vertex and self-energy) amplitudes. However, the presence of gauge-mediated scatterings and inverse decays enhances washout effects, which, if too strong, can erase the produced asymmetry. These effects are incorporated into the Boltzmann equations through additional scattering and washout terms. In the non-holomorphic modular framework, Yukawa couplings transform with negative modular weights in contrast to the holomorphic case where the weights are positive. This distinction alters the structure of the neutrino mass matrix and the decay parameters of $\Sigma_1$, leading to characteristic predictions for the $CP$ asymmetry parameter $\varepsilon_\Sigma$. Importantly, $CP$ violating phases are not free parameters but are intrinsically determined by the modular symmetry making the $CP$ asymmetry prediction highly constrained.


\begin{figure}[t]
    \centering
\begin{tikzpicture}
\begin{feynman}
\vertex at (0,0) (i1);
\vertex at (-2,0) (i2);
\vertex at (2,1) (c);
\vertex at (2,-1) (d);
\vertex at (1,1) (a);
\vertex at (1,-1) (b);
\vertex at (1.5,-1.2) () {\(~\)};

\diagram*{
(i2) -- [fermion, edge label=\(\Sigma_i\)] (i1), (i1) -- [scalar, blue, edge label=\(H\)] (a), (b) -- [fermion, edge label=\({L_i}\)] (i1),
};
\end{feynman}
\end{tikzpicture}\vspace{1cm}
\begin{tikzpicture}
\begin{feynman}
\vertex at (0,0) (i1);
\vertex at (-2,0) (i2);
\vertex at (2,1) (c);
\vertex at (2,-1) (d);
\vertex at (1,1) (a);
\vertex at (1,-1) (b);
\vertex at (1.5,-1.2) () {\(H\)};
\diagram*{
(i2) -- [fermion, edge label=\(\Sigma_i\)] (i1), (i1) -- [scalar, blue, edge label=\(H\)] (a), (b) -- [fermion, edge label=\({L_j}\)] (i1),
(a) --[plain, edge label=\(\Sigma_j\)] (b), (a) -- [fermion, edge label=\({L_i}\)] (c), (b) -- [scalar,blue] (d),
};
\end{feynman}
\end{tikzpicture}\hspace{1cm}
   \caption{Feynman diagrams contributing to the $CP$ asymmetry at tree and one loop-level.}
    \label{feyn-diagram}
\end{figure}
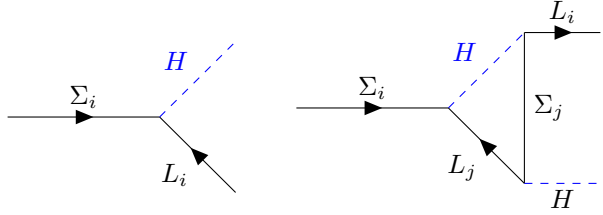

\noindent The interplay of gauge interactions and fixed modular Yukawa structures determines whether the generated lepton asymmetry survives washout and contributes to the baryon asymmetry. In thermal leptogenesis with a hierarchical spectrum of heavy fermion triplets, obtaining the observed baryon asymmetry imposes a lower bound on the mass of the lightest triplet, analogous to the Davidson--Ibarra bound in the Type-I seesaw: $M_{\Sigma_1} \gtrsim 10^{9}~\text{GeV}$~\cite{Davidson:2002qv}. However, the presence of $SU(2)_L$ gauge interactions further tightens this bound as these interactions keep the triplets in thermal equilibrium longer delaying their decoupling \cite{Vatsyayan:2022rth}. Additionally, the reheating temperature after inflation must satisfy $T_{\text{RH}} \gtrsim M_{\Sigma_1}$ to ensure thermal production of triplets. For triplet masses in the TeV range, gauge scatterings remain efficient well below the decoupling scale, erasing any generated asymmetry unless resonant enhancement or non-thermal production is invoked. However, in our model, the Yukawa structure and $CP$ phases are fixed by the non-holomorphic modular symmetry, ruling out such resonant enhancements. Consequently, a high leptogenesis scale, $M_{\Sigma_1} \sim 10^{12}~\text{GeV}$, emerges naturally in our construction. The $CP$ asymmetry can be expressed as\cite{Fong:2012buy} 

\begin{equation}
    \epsilon_{\Sigma i} = \frac{\Gamma(\Sigma_i \to L H) - \Gamma(\bar\Sigma_i \to \bar L \bar H)}{\Gamma(\Sigma_i \to L H) + \Gamma(\bar\Sigma_i \to \bar L \bar H)},
\end{equation}
where $\Gamma(i\rightarrow f)$ is the average decay rate\footnote{In this expression, we have neglected the Pauli blocking and Boltzmann enhancement statistical factors and assume that the initial particle $i$ follows the Maxwell-Boltzmann distribution $i.e. $ $f_i = e^{-\frac{E_i}{T}}$. For detailed analysis that includes the quantum statistical corrections, see refs \cite{Covi:1997dr,Giudice:2003jh}.}

\begin{equation}
\Gamma(i \rightarrow f) \equiv \int \frac{d^3 p_i}{(2\pi)^3 2E_i} \, \frac{d^3 p_f}{(2\pi)^3 2E_f} \, (2\pi)^4 \delta^{(4)}(p_i - p_f) \, |\mathcal{A}(i \rightarrow f)|^2 \, e^{-E_i/T}.
\end{equation}

\begin{figure}
\centering    
  
  \includegraphics[width=0.45\linewidth]{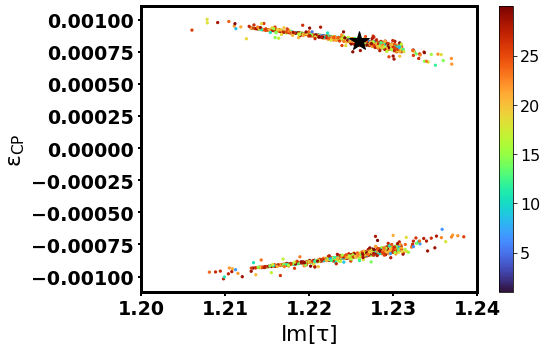}
 
\caption{Variation in the $CP$ asymmetry as function of $\text{Im}[\tau]$.}
\label{cpasy}
\end{figure}

\noindent Here, $\mathcal{A}(i \rightarrow f)$ is the decay amplitude given as 
\begin{equation}
    \epsilon_{\Sigma i} = \frac{|\mathcal{A_o}(\Sigma_i \to L H)|^2 - |\mathcal{A_o}(\bar\Sigma_i \to \bar L \bar H)|^2}{|\mathcal{A_o}(\Sigma_i \to L H)|^2 + |\mathcal{A_o}(\bar\Sigma_i \to \bar L \bar H)|^2},
\label{eqn51}
\end{equation}
where $\mathcal{A_o}$ is the decay amplitude at zero temperature. Eqn. (\ref{eqn51}) receives no contribution at tree level but is generated at the one-loop level via the interference between the tree-level amplitude and the one-loop diagrams as shown in Fig \ref{feyn-diagram}. At leading order, this interference gives rise to the $CP$ asymmetry

\begin{equation}
\varepsilon_{i} = 
\frac{1}{8\pi} \frac{1}{(\tilde{Y}^\dagger \tilde{Y})_{ii}} 
\sum_{j \ne i} 
\mathrm{Im} \left[ (\tilde{Y}^\dagger \tilde{Y})_{ji} \tilde{Y}_{\alpha i} \tilde{Y}^*_{\alpha j} \right] 
\, g\left( \frac{M_{\Sigma_j}^2}{M_{\Sigma_i}^2} \right)
\label{eq:flavoured_cp_asym},
\end{equation}

\noindent
where $\tilde{Y}$ = $Y U_L U_R$ with Y = $\frac{M_D}{v}$ (see Eqn. (\ref{mdyukawa}))\cite{Mishra:2020gxg} and $g$ is the loop function defined as 

\begin{equation}
g(x_j) = \sqrt{x_j} \left[ \frac{1}{1 - x_j} + 1 - (1 + x_j) \ln\left( \frac{1 + x_j}{x_j} \right) \right],
\label{eq:loop_function}
\end{equation}

\noindent
 where, $x_j = \frac{M_{\Sigma_j}}{M_{\Sigma_1}}$. The first term in Eqn.~\eqref{eq:flavoured_cp_asym} arises from lepton-number-violating wave and vertex diagrams. So the $CP$ asymmetry produced by the lightest right-handed neutrino can be calculated as 
\begin{equation}
    \epsilon_i = -\sum_{j = 2,3} \frac{3}{2}\frac{M_{\Sigma_i}}{M_{\Sigma_j}} \frac{\Gamma_j}{M_{\Sigma_j}} I_j \frac{V_j - 2 S_j}{3},
    \end{equation}
where the asymmetry part includes the contribution from all three components of the fermion triplet. However, only the decay of the lightest right-handed neutrino will contribute to the leptogenesis process as the asymmetry generated by the heavier right-handed neutrinos can be washed out by the $\Sigma_1$ mediated interactions until they drop out of equilibrium. 
Here

\begin{equation}
    I_j = \frac{Im[(\tilde{Y}^\dagger \tilde{Y})_{ij}^2]}{(\tilde{Y}^\dagger \tilde{Y})_{ii}(\tilde{Y}^\dagger \tilde{Y})_{jj}},
\end{equation}
and $V_j$ and $S_j$ are the loop factors associated with the vertex and self-energy corrections, respectively, given by 

\begin{equation}
    V_j = \frac{M_{\Sigma_j}^2 (M_{\Sigma_j}^2 - M_{\Sigma_i}^2)}
{(M_{\Sigma_j}^2 - M_{\Sigma_i}^2)^2 + M_{\Sigma_i}^2 \Gamma_{\Sigma_j}^2},
\end{equation}
and
\begin{equation}
    S_j = 2 \frac{M_{\Sigma_j}^2} {M_{\Sigma_i}^2}( \left( 1 + \frac{M_{\Sigma_j}^2}{M_{\Sigma_i}^2} \right)  
\ln \left( 1 + \frac{M_{\Sigma_j}^2}{M_{\Sigma_i}^2} \right) - 1).
\end{equation}
In the hierarchical limits($M_{\Sigma _1}< M_{\Sigma_2} < M_{\Sigma _3} $), the loop factors reduce to unity \cite{Hambye:2012fh, Albright:2003xb}. Further, $\Gamma_j$ represents the decay width of the triplet fermion and can be expressed as 

\begin{equation}
\Gamma_{\Sigma_j} = \left( \frac{|(\tilde{Y}^\dagger \tilde{Y})_{jj}|}{8\pi} \right) M_{\Sigma_j}.
\end{equation}
\noindent The Boltzmann equation (BEs) plays an important role in tracking the evolution of lepton asymmetry as the universe gradually cools down over time. The relevant coupled BEs are given by
\begin{equation}
s \textbf{H} z \frac{dY_\Sigma}{dz} = -\gamma_D \left( \frac{Y_\Sigma}{Y^{\text{eq}}_\Sigma} - 1 \right) - 2 \gamma_A \left( \frac{Y_\Sigma^2}{(Y^{\text{eq}}_\Sigma)^2} - 1 \right),
\label{ysigma}
\end{equation}

\begin{equation}
s \textbf{H} z \frac{dY_{B-L}}{dz} = -\gamma_D \, \epsilon_\Sigma \left( \frac{Y_\Sigma}{Y^{\text{eq}}_\Sigma} - 1 \right) - \frac{Y_{B-L}}{Y^{\text{eq}}_l} \left( \frac{\gamma_D}{2} + \gamma^{\text{sub}}_\Sigma \right),
\end{equation}
where, $Y_{\Sigma a}$ = $\frac{n_a(z)}{s(z)}$ represents the number density of particle species $a$. $s(z)$ is the entropy density given as $s(z)= 0.44 g_* T^3$, $z$ is a dimensionless parameter varying inversely with the temperature of the universe, given as $z \equiv \frac{M_{\Sigma_i}}{T}$. $\textbf{H}$ represents the Hubble expansion rate, given as $\textbf{H} = 1.66 \, \frac{\sqrt{g_*} \, T^2}{M_{\text{Pl}}}$. $\gamma$ denotes the reaction density of processes under consideration, and `D' denotes the decay processes and given as 
\begin{equation}
    \gamma _D (z) = s(z) Y_\Sigma^{eq} \Gamma _{\Sigma} \frac{K_1(z)}{K_2(z)},
\end{equation}

\noindent where $K_1(z)$ and $K_2(z)$ are the modified Bessel functions and $\gamma _A$ in Eqn. (\ref{ysigma}) denotes the gauge annihilation processes and represented as 
\begin{equation}
\gamma _A(z) = \frac{M_{\Sigma_1} T^3}{32\pi^3} e^{-2z}  
\left[
\frac{111 g^4}{8\pi} + \frac{3}{2z} \left( \frac{111 g^4}{8\pi} + \frac{51 g^4}{16\pi} \right) + \mathcal{O}\left(\frac{1}{z^2}\right)
\right],
\end{equation}

\noindent where, $g$ is the typical gauge coupling. The $\gamma^{\text{sub}}$ is for wahsout effects by $\Delta$L = 2 processes. The equilibrium yields are given by

\begin{equation}
    Y_\Sigma^{eq} = \frac{135 g_\Sigma}{16 \pi^4 g_*} z^2 K_2(z), \quad Y_l^{eq} = \frac{3}{4} \frac{45 \zeta(3) g_l}{2 \pi^4 g_*},
\end{equation}

\noindent where $g_l = 2$, $g_\Sigma = 2$ and $g_* = 106.75$.

\noindent If the decays of the heavy right-handed neutrinos proceed out of equilibrium, that is, if they are not significantly faster than the Hubble expansion rate $\textbf{H}(T)$ of the Universe at temperature $T = M_{\Sigma_i}$, a net lepton asymmetry can be preserved. The departure from thermal equilibrium is typically quantified by the decay parameter ($k$), which is defined in terms of the decay rate $\Gamma_\Sigma$ and read as follows
\begin{equation}
    k = \frac{\Gamma_\Sigma}{\textbf{H}(T = M_{\Sigma_1})} = \frac{\tilde{m_i}}{m^*},
\end{equation}
where the effective neutrino mass is given as 
\begin{equation}
    \tilde{m_i} = \frac{(\tilde{Y}^\dagger \tilde{Y})_{ii}v^2}{M_{\Sigma i}},
\end{equation}
and $m^* = \left. 8\pi\frac{ v^2}{M_{\Sigma_1}} \right|_{T = M_{\Sigma_1}}$. We distinguish three regimes based on the value of the $k$ as $k \ll 1$, $k \simeq 1$, and $k \gg 1$, corresponding to the weak, intermediate, and strong washout regimes, respectively. Using the best-fit values of the free parameters in the model, we find that the system consistently lies in the strong washout regime. Consequently, the contributions from $\Delta L = 1$ and $\Delta L = 2$ scattering processes can be safely neglected as mentioned in \cite{Davidson:2008bu, Marciano:2024nwm}. Therefore, in this analysis, we consider only decays and inverse decays. The produced $B-L$ asymmetry converted to the 
\begin{equation}
Y_{B} = c_s Y_{B-L},
\qquad 
c_s = \frac{8N_f + 4N_H}{22N_f + 13N_H}.
\end{equation}
For Type-III seesaw with three heavy fermion\footnote{We consider three generation of fermion triplet and one Higgs. Therefore, $N_f=3$ and $N_H=1$ in the sphaleron conversion factor.}: $N_f = 3, N_H = 1$,
\begin{equation}
c_s = \frac{28}{79},
\qquad 
Y_{B} = \frac{28}{79} Y_{B-L}.
\end{equation}

\noindent Fig.~\ref{cpasy} shows the dependence of the $CP$ asymmetry parameter, $\varepsilon_{CP}$, on the imaginary component of the modular parameter, $\mathrm{Im}[\tau]$. The numerical solutions exhibit two distinct branches: 
one with positive values and one with negative values of $\varepsilon_{CP}$ both of magnitude $|\varepsilon_{CP}| \sim 10^{-3}$. The appearance of two branches originates from discrete transformations of the modular forms entering the Yukawa couplings, which result in equivalent neutrino mass spectra but opposite $CP$ phases.

\begin{table}[t]
\centering
\begin{tabular}{c c c c l}
\hline
Benchmark Point (BP) & $M_{\Sigma_1}$ (GeV) & $\chi^2$ & $\epsilon_{CP}$ & Complex modulus \boldmath$\tau$ \\
\hline
BP1 & $1.16 \times10^{12}$ & 1.03 & -0.0008 & $0.20 + 1.22i$ \\
BP2 & $1.27\times10^{12}$ & 6.48 & 0.0006& $-0.17 + 1.23i$ \\
BP3 & $1.04 \times10^{12}$ & 27.6 & 0.0010&$-0.25 + 1.21 i$ \\
\hline
\end{tabular}
\caption{Benchmark points used to generate the lepton asymmetry in the model.}
\label{bps}
\end{table}

\begin{figure}[t]
    \begin{subfigure}{0.45\textwidth}
  
  \includegraphics[width=1\linewidth]{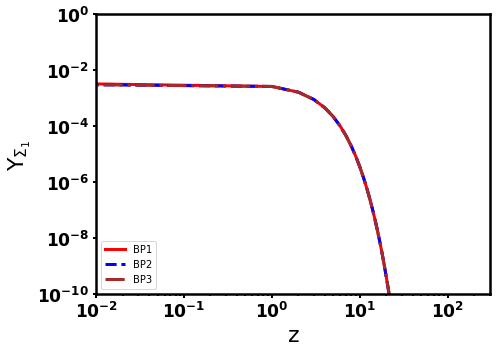}
  \caption{}
  \label{fig:sigma1}
\end{subfigure}%
\hfill
\begin{subfigure}{0.45\textwidth}
  
  \includegraphics[width=1\linewidth]{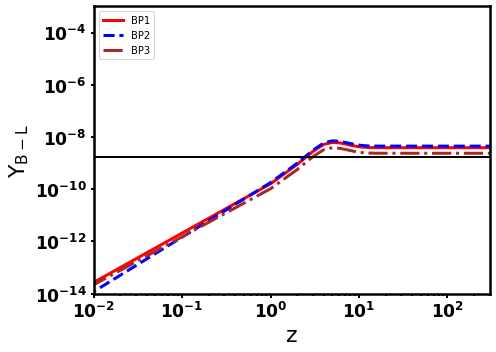}
  \caption{}
  \label{fig:bl}
\end{subfigure}%
    
\caption{Evolution of the comoving number density of the lightest fermion triplet $Y_{\Sigma_1}$ (\textbf{left}) and the $B-L$ asymmetry $Y_{B-L}$ (\textbf{right}) as functions of $z = M_{\Sigma_1}/T$ for three benchmark points 
(BP1, BP2, BP3) given in Table \ref{bps}.}
\label{b-l}
\end{figure}

\noindent The evolution of the comoving number density of the lightest fermion triplet, $Y_{\Sigma_1}$, and the $B-L$ asymmetry, $Y_{B-L}$, as functions of the dimensionless variable $z = M_{\Sigma_1}/T$ is shown in Fig.~\ref{b-l}(\subref{fig:sigma1}) and Fig.~\ref{b-l}(\subref{fig:bl}), respectively. For all benchmark points (BP1, BP2, BP3) shown in Table \ref{bps}, the thermal evolution exhibits similar qualitative behavior, demonstrating the robustness of the results against moderate variations in the model parameters. At high temperatures ($z \ll 1$), the triplet yield remains equal to its equilibrium value, $Y_{\Sigma_1} \simeq Y^{\rm eq}_{\Sigma_1}$, due to the dominance of gauge-mediated scattering processes ($\gamma_A$). These processes are efficient enough to maintain equilibrium until $z \sim \mathcal{O}(1)$. As the temperature drops ($z \gtrsim 1$), the gauge scattering rate falls rapidly below the Hubble expansion rate, and the triplet population begins to depart from equilibrium. Subsequently, the number density of $\Sigma_1$ is exponentially suppressed for $z \gg 10$, consistent with its expected non-relativistic freeze-out behavior. The $B-L$ yield, $Y_{B-L}$, starts from zero and grows in the temperature range $z \sim 1-10$, where the decays of $\Sigma_1$ into lepton-Higgs final states dominate over inverse decays and washout effects. The $CP$ violating nature of these decays, quantified by the asymmetry parameter $\varepsilon_{CP}$, results in a net lepton number generation. For $z \gg 10$, both inverse decays and gauge scatterings become negligible, causing washout processes to freeze out and the $B-L$ asymmetry to reach a constant value.

\begin{figure}[t]
    \begin{subfigure}{0.45\textwidth}
  
  \includegraphics[width=1\linewidth]{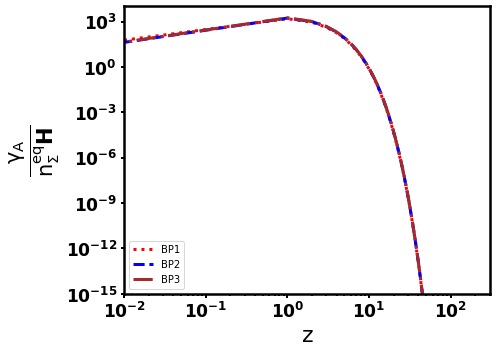}
  \caption{}
  \label{figg1}
\end{subfigure}%
\hfill
\begin{subfigure}{0.45\textwidth}
  
  \includegraphics[width=1\linewidth]{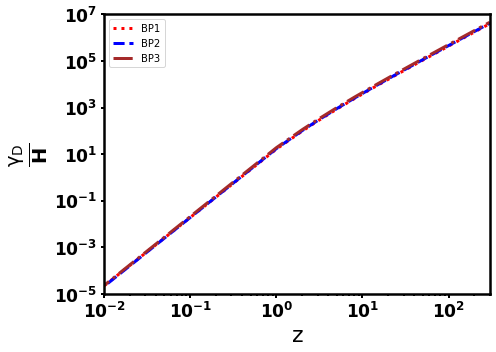}
  \caption{}
  \label{figg2}
\end{subfigure}%
\hfill
 \centering
 \begin{subfigure}{0.45\textwidth}
  
  \includegraphics[width=1\linewidth]{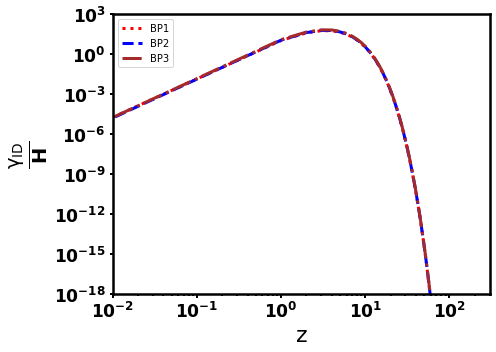}
  \caption{}
  \label{figg3}
\end{subfigure}%
\hfill
\centering
\begin{subfigure}{0.45\textwidth}
    \includegraphics[width=1\linewidth]{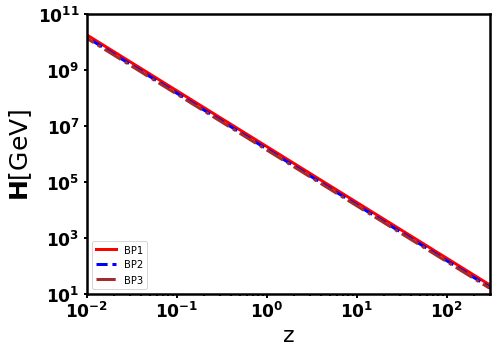}
    \caption{}
    \label{figg4}
\end{subfigure}

\caption{
Normalized reaction densities and Hubble expansion rate $\textbf{H}$ as functions of $z=M_{\Sigma_1}/T$ for three benchmark points (BP1, BP2, BP3): 
(a) $\gamma_A/(n^{eq}_{\Sigma}\textbf{H})$, 
(b) $\gamma_D/\textbf{H}$, 
(c) $\gamma_{ID}/\textbf{H}$, 
(d) $\textbf{H}$.}

\label{figgauge}
\end{figure}

\noindent The thermal evolution of the relevant processes governing the dynamics of the fermion triplet $\Sigma$ has been analyzed by comparing the gauge-mediated scattering $(\gamma_{A})$, decay $(\gamma_{D})$ and inverse decay $(\gamma_{ID})$ rates with $z$ depicted in Fig. \ref{figgauge}(\subref{figg1}), \ref{figgauge}(\subref{figg2}) and \ref{figgauge}(\subref{figg3}), respectively. All reaction densities are normalized to the Hubble rate to capture the relative strength of the individual processes as a function of the dimensionless parameter $z = M_{\Sigma}/T$. For all benchmark points (BP1, BP2, BP3), the Hubble rate exhibits the expected radiation-dominated scaling, $\textbf{H} \propto z^{-2}$, reflecting its independence from the triplet sector parameters as depicted in \ref{figgauge}(\subref{figg4}). In the high-temperature regime ($z \lesssim 1$), gauge scatterings $\Sigma\Sigma \leftrightarrow f\bar{f}$ and $\Sigma f \leftrightarrow \Phi f'$ are dominant, with $\gamma_{A}/(n^{\mathrm{eq}}_{\Sigma} \textbf{H}) \gg 1$, ensuring that the triplet population remains in thermal equilibrium. As the Universe cools and $z$ approaches $\mathcal{O}(1-10)$, gauge scattering rates decrease rapidly due to the Boltzmann suppression of the equilibrium triplet number density, leading to the eventual decoupling of these processes (shown in Fig. \ref{figgauge}(\subref{figg1})). The decay rate, normalized to the Hubble parameter $\gamma_{D}/\textbf{H}$, is suppressed in the relativistic regime ($z \ll 1$) but increases monotonically, becoming comparable to or exceeding the Hubble rate for $z \gtrsim 1$. As shown in Fig. \ref{figgauge}(\subref{figg2}), this behavior ensures that decays dominate over the cosmic expansion at late times, enabling the out-of-equilibrium condition required for leptogenesis. The inverse decay rate, $\gamma_{ID}/\textbf{H}$, follows a similar trend but peaks near $z \sim \mathcal{O}(1-10)$ and falls exponentially at larger $z$, indicating that washout effects are most significant around the transition region and become negligible once the triplet abundance is suppressed (shown in Fig. \ref{figgauge}(\subref{figg3})). The combined behavior of these rates confirms the canonical picture of triplet leptogenesis: (i) strong gauge interactions enforce thermal equilibrium at early times, (ii) the weakening of these scatterings allows decays to dominate, and (iii) inverse decays cease to wash out the generated asymmetry in the late, non-relativistic regime. The near-identical behavior of all three benchmark points indicates that the qualitative thermal history is insensitive to modest variations in model parameters, implying a robust parameter space for successful leptogenesis. Further, the recent developments in Cosmological Collider physics provide a novel probe of high-scale leptogenesis scenarios \cite{Cui:2021iie,Lu:2019tjj,Dvali:2003em,Kofman:2003nx,Suyama:2007bg,Ichikawa:2008ne}. During inflation, heavy fields can leave characteristic oscillatory features in primordial fluctuations, accessible through future CMB, large-scale structure, and 21 cm line surveys. In particular, the Higgs boson, which directly participates in leptogenesis via Yukawa couplings, can mediate distinctive patterns in the primordial bispectrum, sensitive to lepton-number-violating interactions and $CP$ violating phases. In this framework, these $CP$ phases are not free parameters but are determined by the non-holomorphic modular symmetry, leading to a highly predictive structure of the leptogenesis sector. This implies that any cosmological signature associated with leptogenesis, such as the oscillatory shape of the Higgs bispectrum, would be tightly correlated with the fixed $CP$ violating phases. Thus, precision cosmological measurements of non-Gaussianity could provide an indirect probe of both the high leptogenesis scale and the modular origin of $CP$ violation in our scenario.

\section{Conclusions}
\label{section6}
We have proposed Type-III seesaw model for neutrino mass generation and successful baryogenesis in the non-holomorphic modular symmetry framework. The lack of experimental evidences for the low-energy SUSY and exploration incorporating non-SUSY scenarios are claiming. In comparison to the holomorphic modular symmetry, the Yukawa couplings are given by the polyharmonic Maaß forms where the assumption of holomorphicity is replaced by the Laplacian condition, while modularity remains preserved.  We have incorporated SU(2) fermion triplet($\Sigma_i$), which is transforming as triplet under $A_4$ with modular weight $-2$. 
We have performed the chi-square analysis to test the model against the observed neutrino oscillation data. The model predicts the neutrino oscillation data, lepton masses ratio and BAU with NH only, whereas the IH is ruled out by neutrino oscillation data at 3$\sigma$ level. 
The minimum value of chi-square is 1.03 and the parameters corresponding to this value are $\sin^2\theta_{13} = 0.022$, $\sin^2\theta_{12} = 0.30$, $\sin^2\theta_{23} = 0.47$, $\Delta m_{21}^2 = 0.000075~ \text{eV}^2$, $\Delta m_{31}^2 = 0.0025~\text{eV}^2$, $\frac{me}{m_\mu} = 0.0047$, $\frac{m_\mu}{m_\tau} = 0.058$, $\delta _{CP} = 22.5^o$, $\alpha_{21} = -73.3^o $, $\alpha_{31}= 3.99^o$, $\tau=0.20+1.22 i$, $m_1 = 0.027$ eV, $m_{\beta \beta} = 0.022$ eV and $M_{\Sigma_1} = 1.04 \times 10^{12}$ GeV. The model favors first octant of $\theta_{23}$, which shows compatibility with the joint result of T2K and NO$\nu$A results. Further, we have computed the effective Majorana mass, which is with in the sensitivity of nEXO experiment. Additionally, the sum of neutrino masses lies in the range (0.104-0.116) eV in line with the cosmological bound.

\noindent Further, non-holomorphic modular Type-III seesaw model, studied in the present work, but, also provides a natural 
framework for generating the baryon asymmetry of the Universe via thermal leptogenesis. The $CP$ phases, fixed by the modular symmetry rather than arbitrary choices yield a robust $CP$ asymmetry of order $\epsilon_{CP}\sim 10^{-3}$. Combined with the 
strong gauge-mediated washout of fermion triplets, this necessitates a leptogenesis scale of $\mathcal{O}(10^{12}\,\mathrm{GeV})$ ensuring compatibility with both the Davidson-Ibarra bound and the thermal history of the Universe. While direct experimental tests of such a high scale remain challenging, future cosmological collider probes and precision measurements of primordial non-Gaussianity may provide indirect signatures of the $CP$ phases and lepton-number-violating interactions predicted by this framework. Future endevours are open to explore avenues for lowering the leptogenesis scale.

\section*{Acknowledgments}
LS gratefully acknowledges the financial support from the Council of Scientific and Industrial Research (CSIR), Government of India, under the grant No. 09/1196(18553)/2024-EMR-I.
\appendix
\section*{Appendix}

\subsection*{Modular Group $\Gamma_3 \cong A_4$ and Polyharmonic Maaß forms of level N = 3}



For two $A_4$ triplets $\alpha = (\alpha_1, \alpha_2, \alpha_3)^T$ and $\beta = (\beta_1, \beta_2, \beta_3)^T$, the tensor product 
decomposes as $3 \otimes 3 = 1 \oplus 1' \oplus 1'' \oplus 3_S \oplus 3_A$, where $3_S$ and $3_A$ are the symmetric and antisymmetric parts respectively and the contraction rules for forming these irreducible representation are given as 

\begin{align}
\begin{pmatrix}
\alpha_1 \\
\alpha_2 \\
\alpha_3
\end{pmatrix}_3
\otimes
\begin{pmatrix}
\beta_1 \\
\beta_2 \\
\beta_3
\end{pmatrix}_3
&=
(\alpha_1 \beta_1 + \alpha_2 \beta_3 + \alpha_3 \beta_2)_{1} 
\oplus (\alpha_3 \beta_3 + \alpha_1 \beta_2 + \alpha_2 \beta_1)_{1'} 
\oplus (\alpha_2 \beta_2 + \alpha_1 \beta_3 + \alpha_3 \beta_1)_{1''} \nonumber \\ 
&\quad \oplus
\begin{pmatrix}
2\alpha_1 \beta_1 - \alpha_2 \beta_3 - \alpha_3 \beta_2 \\
2\alpha_3 \beta_3 - \alpha_1 \beta_2 - \alpha_2 \beta_1 \\
2\alpha_2 \beta_2 - \alpha_1 \beta_3 - \alpha_3 \beta_1
\end{pmatrix}_{3_{sym}}
\oplus
\begin{pmatrix}
\alpha_2 \beta_3 - \alpha_3 \beta_2 \\
\alpha_1 \beta_2 - \alpha_2 \beta_1 \\
\alpha_3 \beta_1 - \alpha_1 \beta_3
\end{pmatrix}_{3_{asym}}.
\end{align}
There are three modular forms of weight 2 and level 3, which together form an irreducible triplet representation 3 of the group $ A_4 $. These modular forms can be written as
\begin{equation}
Y^{(2)}_3 =
\begin{pmatrix}
Y_1(\tau) \\
Y_2(\tau) \\
Y_3(\tau)
\end{pmatrix},
\end{equation}
with the $q$-expansions

\begin{align}
Y_1(\tau) &= 1 + 12q + 36q^2 + 12q^3 + 84q^4 + 72q^5 + \dots, \\
Y_2(\tau) &= -6q^{1/3} (1 + 7q + 8q^2 + 18q^3 + 14q^4 + \dots), \\
Y_3(\tau) &= -18q^{2/3} (1 + 2q + 5q^2 + 4q^3 + 8q^4 + \dots).
\end{align}

\noindent For integer weights $ k > 2 $, the polyharmonic Maaß forms at level 3 are the same as the corresponding modular forms. In contrast, at weight 2 and level 3, the space of polyharmonic Maaß forms is larger and includes additional non-holomorphic functions. This space contains the modified Eisenstein series $ E^{(b)}_2(\tau) $, which transforms as a trivial singlet under the $ A_4 $ symmetry, and the modular triplet $ Y^{(2)}_3 $, which transforms as an irreducible triplet of $ A_4 $. As a result, the general structure of the weight 2 polyharmonic Maaß form can be written as a combination of these components
\begin{equation}
    Y^{(2)}_1 \equiv E^{(b)}_2(\tau).
\end{equation}
The presence of the modular form $ Y^{(2)}_3 $ indicates that, apart from the trivial solution $ Y^{(0)}_1 = 1 $, there also exists a non-trivial polyharmonic Maaß form $Y^{(0)}_3$ with weight 0 and the associated master equation has the following form

\begin{align}
    Y^{(0)}_{3,1} &= y - \frac{3 e^{-4\pi y}}{\pi q} - \frac{9 e^{-8\pi y}}{2\pi q^2} 
    - \frac{e^{-12\pi y}}{\pi q^3} - \frac{21 e^{-16\pi y}}{4\pi q^4} 
    - \frac{18 e^{-20\pi y}}{5\pi q^5} - \frac{3 e^{-24\pi y}}{2\pi q^6} + \dots \nonumber \\
    &\quad - \frac{9 \log 3}{4\pi} - \frac{3q}{\pi} - \frac{9q^2}{2\pi} - \frac{q^3}{\pi} 
    - \frac{21q^4}{4\pi} - \frac{18q^5}{5\pi} - \frac{3q^6}{2\pi} + \dots, 
\end{align}

\begin{align}
    Y^{(0)}_{3,2} &= \frac{27 q^{1/3} e^{\pi y /3}}{\pi} \left( \frac{e^{-3\pi y}}{ 4q} 
    + \frac{e^{-7\pi y}} {5q^2} + \frac{5 e^{-11\pi y}} {16q^3} + \frac{2 e^{-15\pi y}} {11q^4} + \frac{2 e^{-19\pi y}} {7q^5 }
    + \frac{4 e^{-23\pi y}} {17q^6} + \dots \right) \nonumber \\
    &\quad + \frac{9 q^{1/3}}{2\pi} \left( 1 + \frac{7q}{4} + \frac{8q^2} {7} + \frac{9q^3} {5} + \frac{14q^4} {13} 
    + \frac{31q^5} {16} + \frac{20q^6} {19} + \dots \right),
\end{align}

\begin{align}
    Y^{(0)}_{3,3} &= \frac{9 q^{2/3} e^{2\pi y /3}}{2\pi} \left( \frac{e^{-2\pi y}}{ q} 
    + \frac{7 e^{-6\pi y}} {4q^2} + \frac{8 e^{-10\pi y}} {7q^3} + \frac{9 e^{-14\pi y}} {5q^4} 
    + \frac{14 e^{-18\pi y}} {13q^5} + \frac{31 e^{-22\pi y}} {16q^6} + \dots \right) \nonumber \\
    &\quad + \frac{27q^{2/3}}{\pi} \left( \frac{1}{4} + \frac{q} {5} + \frac{5q^2} {16} + \frac{2q^3} {11} 
    + \frac{2q^4} {7} + \frac{9q^5} {17} + \frac{21q^6} {20} + \dots \right).
\end{align}
The weight $-2$ polyharmonic Maaß forms of level 3 can be constructed from the weight 4 modular forms $Y^{(4)}_1$, $Y^{(4)}_{1'}$, and $Y^{(4)}_3$. It is important to note that $ Y^{(4)}_{1'}$ is a cusp form in $\Gamma(3)$, and therefore, it cannot be lifted to a polyharmonic Maaß form. The corresponding weight $-2$ Yukawa couplings are given below

\begin{align}
    Y^{(-2)}_1(\tau) &= \frac{y^3}{3} - \frac{15\Gamma(3,4\pi y)}{4\pi^3 q} 
    - \frac{135\Gamma(3,8\pi y)}{32\pi^3 q^2} 
    - \frac{35\Gamma(3,12\pi y)}{9\pi^3 q^3} + \dots \nonumber \\
    &\quad - \frac{\pi}{12} \frac{\zeta(3)}{\zeta(4)} - \frac{15q}{2\pi^3} 
    - \frac{135q^2}{16\pi^3} - \frac{70q^3}{9\pi^3} 
    - \frac{1095q^4}{128\pi^3} - \frac{189q^5}{25\pi^3} 
    - \frac{35q^6}{4\pi^3} + \dots
\end{align}

\begin{align}
    Y^{(-2)}_{3,1}(\tau) &= \frac{y^3}{3} + \frac{21\Gamma(3,4\pi y)}{16\pi^3 q} 
    + \frac{189\Gamma(3,8\pi y)}{128\pi^3 q^2} 
    + \frac{169\Gamma(3,12\pi y)}{144\pi^3 q^3} 
    + \frac{1533\Gamma(3,16\pi y)}{1024\pi^3 q^4} + \dots \nonumber \\
    &\quad + \frac{\pi}{40} \frac{\zeta(3)}{\zeta(4)} + \frac{21q}{8\pi^3} 
    + \frac{189q^2}{64\pi^3} + \frac{169q^3}{72\pi^3} 
    + \frac{1533q^4}{512\pi^3} + \frac{1323q^5}{500\pi^3} 
    + \frac{169q^6}{64\pi^3} + \dots
\end{align}

\begin{align}
    Y^{(-2)}_{3,2}(\tau) &= -\frac{729q^{1/3}}{16\pi^3} 
    \left( \frac{\Gamma(3,8\pi y/3)}{16q} 
    + \frac{7\Gamma(3,20\pi y/3)}{125q^2} 
    + \frac{65\Gamma(3,32\pi y/3)}{1024q^3} 
    + \frac{74\Gamma(3,44\pi y/3)}{1331q^4} + \dots \right) \nonumber \\
    &\quad -\frac{81q^{1/3}}{16\pi^3} 
    \left( 1 + \frac{73q}{64} 
    + \frac{344q^2}{343} 
    + \frac{567q^3}{500} 
    + \frac{20198q^4}{2197} 
    + \frac{4681q^5}{4096} + \dots \right),
\end{align}

\begin{align}
    Y^{(-2)}_{3,3}(\tau) &= -\frac{81q^{2/3}}{32\pi^3} 
    \left( \frac{\Gamma(3,4\pi y/3)}{q} 
    + \frac{73\Gamma(3,16\pi y/3)}{64q^2} 
    + \frac{344\Gamma(3,28\pi y/3)}{343q^3} 
    + \frac{567\Gamma(3,40\pi y/3)}{500q^4} + \dots \right) \nonumber \\
    &\quad -\frac{729q^{2/3}}{8\pi^3} 
    \left( \frac{1}{16} 
    + \frac{7q}{125} 
    + \frac{65q^2}{1024} 
    + \frac{74q^3}{1331} + \dots \right).
\end{align}


\begin{thebibliography}{100}



\bibitem{Esteban:2024eli}
I.~Esteban, M.~C.~Gonzalez-Garcia, M.~Maltoni, I.~Martinez-Soler, J.~P.~Pinheiro and T.~Schwetz,
JHEP \textbf{12}, 216 (2024).


\bibitem{Petcov:2017ggy}
S.~T.~Petcov,
Eur. Phys. J. C \textbf{78}, no.9, 709 (2018).

\bibitem{Ding:2024ozt}
G.~J.~Ding and J.~W.~F.~Valle,
Phys. Rept. \textbf{1109}, 1-105 (2025).

\bibitem{Ishimori:2010au}
H.~Ishimori, T.~Kobayashi, H.~Ohki \textit{et al.}
Prog. Theor. Phys. Suppl. \textbf{183}, 1-163 (2010).

\bibitem{Chauhan:2023faf}
G.~Chauhan, P.~S.~B.~Dev \textit{et al.}
Prog. Part. Nucl. Phys. \textbf{138}, 104126 (2024).


\bibitem{Feruglio:2017spp}
F.~Feruglio,
[arXiv:1706.08749 [hep-ph]].

\bibitem{Feruglio:2019ybq}
F.~Feruglio and A.~Romanino,
Rev. Mod. Phys. \textbf{93}, no.1, 015007 (2021).

\bibitem{Ohki:2020bpo}
H.~Ohki, S.~Uemura and R.~Watanabe,
Phys. Rev. D \textbf{102}, no.8, 085008 (2020).

\bibitem{deAdelhartToorop:2011re}
R.~de Adelhart Toorop, F.~Feruglio and C.~Hagedorn,
Nucl. Phys. B \textbf{858}, 437-467 (2012).


\bibitem{ParticleDataGroup:2024cfk}
S.~Navas \textit{et al.} [Particle Data Group],
Phys. Rev. D \textbf{110}, no.3, 030001 (2024).

\bibitem{Cremades:2004wa}
D.~Cremades, L.~E.~Ibanez and F.~Marchesano,
JHEP \textbf{05}, 079 (2004).

\bibitem{Almumin:2021fbk}
Y.~Almumin, M.~C.~Chen \textit{et al.}
JHEP \textbf{05}, 078 (2021).

\bibitem{Qu:2024rns}
B.~Y.~Qu and G.~J.~Ding,
JHEP \textbf{08}, 136 (2024).

\bibitem{Nomura:2024atp}
T.~Nomura and H.~Okada,
Phys. Lett. B \textbf{868}, 139763 (2025).

\bibitem{Nomura:2024vzw}
T.~Nomura, H.~Okada and O.~Popov,
Phys. Lett. B \textbf{860}, 139171 (2025).

\bibitem{Kobayashi:2025hnc}
T.~Kobayashi, H.~Okada and Y.~Orikasa,
[arXiv:2502.12662 [hep-ph]].

\bibitem{Ding:2024inn}
G.~J.~Ding, J.~N.~Lu, S.~T.~Petcov and B.~Y.~Qu,
JHEP \textbf{01}, 191 (2025).

\bibitem{Li:2024svh}
C.~C.~Li, J.~N.~Lu and G.~J.~Ding,
JHEP \textbf{12}, 189 (2024).

\bibitem{Loualidi:2025tgw}
M.~A.~Loualidi, M.~Miskaoui and S.~Nasri,
Phys. Rev. D \textbf{112}, no.1, 015008 (2025).

\bibitem{Zhang:2025dsa}
X.~Zhang and Y.~Reyimuaji,
[arXiv:2507.06945 [hep-ph]].


\bibitem{Nomura:2025raf}
T.~Nomura and H.~Okada,
[arXiv:2506.02639 [hep-ph]].

\bibitem{Nomura:2025ovm}
T.~Nomura, H.~Okada and X.~Y.~Wang,
[arXiv:2504.21404 [hep-ph]].

\bibitem{Okada:2025jjo}
H.~Okada and Y.~Orikasa,
[arXiv:2501.15748 [hep-ph]].

\bibitem{Nomura:2024nwh}
T.~Nomura and H.~Okada,
Phys. Lett. B \textbf{867}, 139618 (2025).

\bibitem{Kumar:2025bfe}
B.~Kumar and M.~K.~Das,
[arXiv:2504.21701 [hep-ph]].


\bibitem{Nomura:2025bph}
T.~Nomura and H.~Okada,
[arXiv:2503.19251 [hep-ph]].

\bibitem{Nomura:2024ctl}
T.~Nomura and H.~Okada,
[arXiv:2409.10912 [hep-ph]].

\bibitem{Foot:1988aq}
R.~Foot, H.~Lew, X.~G.~He and G.~C.~Joshi,
Z. Phys. C \textbf{44}, 441 (1989).




\bibitem{Davidson:2002qv}
S.~Davidson and A.~Ibarra,
Phys. Lett. B \textbf{535}, 25-32 (2002)

\bibitem{Ishimori:2012zz}
H.~Ishimori, T.~Kobayashi, H.~Ohki, et. al.,
Lect. Notes Phys. \textbf{858}, 1-227 (2012).

\bibitem{Mishra:2022egy}
P.~Mishra, M.~K.~Behera, P.~Panda and R.~Mohanta,
Eur. Phys. J. C \textbf{82}, no.12, 1115 (2022).

\bibitem{Kashav:2021zir}
M.~Kashav and S.~Verma,
JHEP \textbf{09}, 100 (2021).

\bibitem{Kashav:2022kpk}
M.~Kashav and S.~Verma,
JCAP \textbf{03}, 010 (2023).



\bibitem{Singh:2024imk}
L.~Singh, M.~Kashav and S.~Verma,
Nucl. Phys. B \textbf{1007}, 116666 (2024).

\bibitem{Kashav:2024lkr}
M.~Kashav and K.~M.~Patel,
Phys. Rev. D \textbf{111}, no.1, 015010 (2025).

\bibitem{Kumar:2023moh}
R.~Kumar, P.~Mishra, M.~K.~Behera, R.~Mohanta and R.~Srivastava,
Phys. Lett. B \textbf{853}, 138635 (2024).

\bibitem{Behera:2020lpd}
M.~K.~Behera, S.~Singirala, S.~Mishra and R.~Mohanta,
J. Phys. G \textbf{49}, no.3, 035002 (2022).

\bibitem{Behera:2020sfe}
M.~K.~Behera, S.~Mishra, S.~Singirala and R.~Mohanta,
Phys. Dark Univ. \textbf{36}, 101027 (2022)



\bibitem{Xing:2007fb}
Z.~z.~Xing, H.~Zhang and S.~Zhou,
Phys. Rev. D \textbf{77}, 113016 (2008).

\bibitem{2852068}
I.~Singh \textit{et al.} [NOvA],
FERMILAB-CONF-24-0792-PPD.

\bibitem{Carabadjac:2024smm}
D.~Carabadjac [T2K],
PoS \textbf{ICHEP2024}, 150 (2025).

\bibitem{DUNE:2020fgq}
B.~Abi \textit{et al.} [DUNE],
Eur. Phys. J. C \textbf{81}, no.4, 322 (2021).

\bibitem{Hyper-Kamiokande:2018ofw}
K.~Abe \textit{et al.} [Hyper-Kamiokande],
[arXiv:1805.04163 [physics.ins-det]].


\bibitem{nEXO:2021ujk}
G.~Adhikari \textit{et al.} [nEXO],
J. Phys. G \textbf{49}, no.1, 015104 (2022).

\bibitem{KATRIN:2022ayy}
M.~Aker \textit{et al.} [KATRIN],
J. Phys. G \textbf{49}, no.10, 100501 (2022).


\bibitem{Project8:2022wqh}
A.~A.~Esfahani \textit{et al.} [Project 8],
[arXiv:2203.07349 [nucl-ex]].


\bibitem{Planck:2018vyg}
N.~Aghanim \textit{et al.} [Planck],
Astron. Astrophys. \textbf{641}, A6 (2020).


\bibitem{Vatsyayan:2022rth}
D.~Vatsyayan and S.~Goswami,
Phys. Rev. D \textbf{107}, no.3, 035014 (2023).

\bibitem{Fong:2012buy}
C.~S.~Fong, E.~Nardi and A.~Riotto,
Adv. High Energy Phys. \textbf{2012}, 158303 (2012).



\bibitem{Covi:1997dr}
L.~Covi, N.~Rius, E.~Roulet and F.~Vissani,
Phys. Rev. D \textbf{57}, 93-99 (1998).

\bibitem{Giudice:2003jh}
G.~F.~Giudice, A.~Notari, M.~Raidal, A.~Riotto and A.~Strumia,
Nucl. Phys. B \textbf{685}, 89-149 (2004).

\bibitem{Mishra:2020gxg}
S.~Mishra,
[arXiv:2008.02095 [hep-ph]].


\bibitem{Hambye:2012fh}
T.~Hambye,
New J. Phys. \textbf{14}, 125014 (2012).

\bibitem{Albright:2003xb}
C.~H.~Albright and S.~M.~Barr,
Phys. Rev. D \textbf{69}, 073010 (2004).


\bibitem{Marciano:2024nwm}
S.~Marciano, D.~Meloni and M.~Parriciatu,
JHEP \textbf{05}, 020 (2024).

\bibitem{Davidson:2008bu}
S.~Davidson, E.~Nardi and Y.~Nir,
Phys. Rept. \textbf{466}, 105-177 (2008).





\bibitem{Cui:2021iie}
Y.~Cui and Z.~Z.~Xianyu,
Phys. Rev. Lett. \textbf{129}, no.11, 111301 (2022).

\bibitem{Lu:2019tjj}
S.~Lu, Y.~Wang and Z.~Z.~Xianyu,
JHEP \textbf{02}, 011 (2020).

\bibitem{Dvali:2003em}
G.~Dvali, A.~Gruzinov and M.~Zaldarriaga,
Phys. Rev. D \textbf{69}, 023505 (2004).

\bibitem{Kofman:2003nx}
L.~Kofman,
[arXiv:astro-ph/0303614 [astro-ph]].

\bibitem{Suyama:2007bg}
T.~Suyama and M.~Yamaguchi,
Phys. Rev. D \textbf{77}, 023505 (2008).

\bibitem{Ichikawa:2008ne}
K.~Ichikawa, T.~Suyama, T.~Takahashi and M.~Yamaguchi,
Phys. Rev. D \textbf{78}, 063545 (2008).





\end{thebibliography}
\end{document}